\documentclass[aps,prx,twocolumn,showkeys,floatfix,nobalancelastpage]{revtex4-2}

\usepackage{graphicx}
\usepackage{subcaption}
\usepackage{dcolumn}
\usepackage{multirow}
\usepackage{physics}
\usepackage{mathtools}
\usepackage{bm}
\usepackage{amssymb}
\usepackage{xcolor}
\usepackage{comment}
\usepackage{ragged2e}
\usepackage{blindtext}
\usepackage{soul}
\usepackage[dvipsnames]{xcolor}
\usepackage{placeins}
\usepackage[hidelinks]{hyperref}
\AtBeginDocument{}

\newcommand{\addAA}[1]{\textcolor{black}{#1}}
\newcommand{\addND}[2]{\textcolor{black}{#1}}

\begin{document}

\preprint{APS/123-QED}

\title{Probing Time Reversal Symmetry Breaking using a Nonlinear Superconducting Ring Resonator}

\author{\small Nicolas Dirnegger\textsuperscript{1}, Marie Wesson\textsuperscript{3}, Arpit Arora\textsuperscript{1,2,*}, 
Ioannis Petrides\textsuperscript{2}, Jonathan B. Curtis\textsuperscript{2,4}, Emily M. Been \textsuperscript{2}, Amir Yacoby\textsuperscript{3}, 
Prineha Narang\textsuperscript{1,2}}
\thanks{\href{mailto:arpit22@ucla.edu}{arpit22@ucla.edu}}
\thanks{\href{mailto:prineha@ucla.edu}{prineha@ucla.edu}}

\affiliation{\small \textsuperscript{1}Electrical and Computer Engineering Department, University of California, Los Angeles, CA 90095, USA}
\affiliation{\small \textsuperscript{2} College of Letters and Science, University of California, Los Angeles, CA 90095, USA}
\affiliation{\small \textsuperscript{3} Department of Physics, Harvard University, MA 02138, USA}
\affiliation{\small \textsuperscript{4} Institute for Theoretical Physics, ETH Zurich, 8093 Zurich, Switzerland}
           
\date{\today} 
           
\begin{abstract}
Time-reversal symmetry breaking (TRSB) has been central to detecting exotic phases of matter. Here, we leverage the circuit electrodynamics capabilities of superconducting devices to propose a novel scheme based on a multimode superconducting ring resonator for sensitive probing of TRSB in quantum materials. A ring resonator enables nonlinear cross-interactions between the modes which act as an built-in amplifiers to be harnessed for enhanced sensing. Using a driven-dissipative model, we explore the nonlinear dynamics of a two-mode superconducting circuit with self- and cross-Kerr nonlinearities under conditions near the bifurcation threshold. By mapping the optimal parameter regimes, we show that even when the photon occupation numbers are subjected to different initial conditions, they can be driven into a symmetric configuration which is broken even with weak TRSB.  Through full quantum analysis we demonstrate that the Kerr-nonlinear interactions up-convert the magnetic effects of material-resonator hybrid system, enhancing the probing of TRSB. Our findings highlight the utility of superconducting microwave resonators outside of quantum information processing, as a tool for probing exotic states of matter. 
\end{abstract}

\maketitle

\section{Introduction} \label{section:Introduction}
Quantum phases of matter reflect the underlying symmetry of their order parameter. 
Particularly, time reversal symmetry broken (TRSB) orders have been shown to play important roles in many exotic phenomena including topology ~\cite{nagaosa2010anomalous,kane2005z,PhysRevB.106.165130}, magnetism ~\cite{machida2010time,aaw3780,aay5533}, and unconventional superconductivity ~\cite{curtis2022proximity,kallin2016chiral,read2000paired,ghosh2020recent}. 
TRSB manifests in nonreciprocal responses of optoelectronics, superconductors, and collective-mode dynamics, thus playing a crucial role in device operations and uncovering previously inaccessible properties of quantum materials ~\cite{nagaosa2024nonreciprocal,bottcher2024circuit,arp2024intervalley}. \addND{However, TRSB phases can be subtle and sometimes require sensitive probes.},

\begin{figure}
    \centering
    \includegraphics[width=\linewidth]{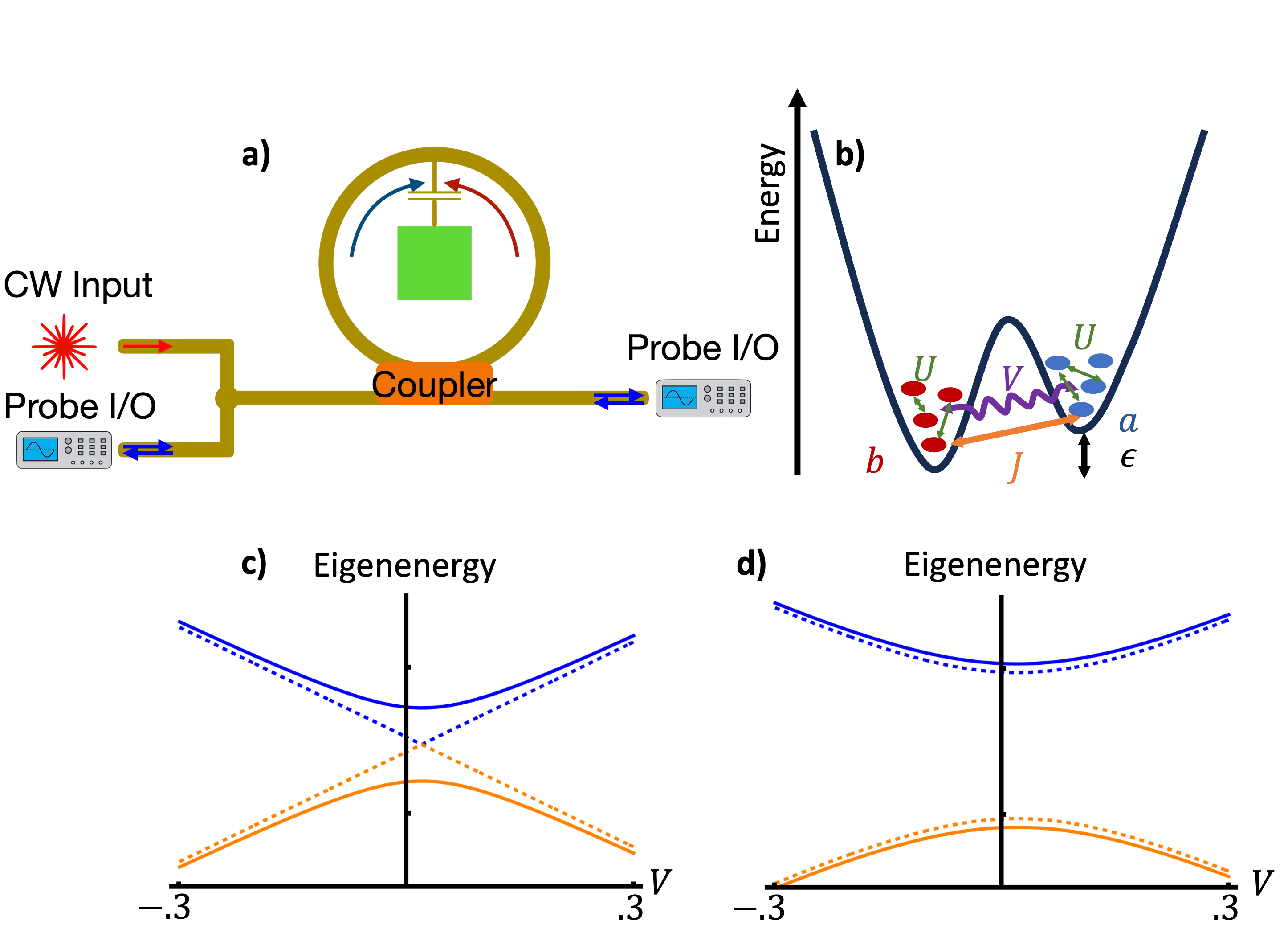}
    \caption{\justifying 
    (a) Schematic of a superconducting ring resonator with counter propagating modes $a$ and $b$.
    The ring resonator is coupled to CW pump via a coupler (depicted in orange) with an external coupling rate $\kappa$, a dissipation rate of modes to the environment $\gamma$, and probed from both sides. The sample material to be probed is indicated with the green square. 
    (b) The two modes $a$ and $b$ as two sites of a Hubbard model with self-Kerr $U$, cross-Kerr $V$, linear exchange interaction strength $J$, and natural splitting between the modes, $\epsilon$.
    (c,d) Energy eigenvalues as a function of $V$ for the model described in Fig. 1b and Eq.~(\ref{eq:RotHam}) without drive.  
    The remaining parameters were chosen to be $U=0.6$, $\text{Re}[J]=0.1$ and $\omega_o = 4.95$.}
    \label{fig:SystemSetup}
\end{figure}

Circuit quantum electrodynamics (cQED) architectures~\cite{cavity_optomechanics,circuit_qed} are emerging as a universal interface for sensing plartforms given their versatility.
Research on non-Hermitian sensors \cite{el2018non} and engineered critical dynamics near phase transitions in cQED lattices \cite{fitzpatrick2017observation}, as well as progress in quantum error correction protocols \cite{ofek2016extending,leghtas2016}, demonstrate that exploiting tunable nonlinearities can substantially boost detection sensitivity and robustness. 
In this work, we propose a novel scheme to probe TRSB in quantum materials that leverages the cross-Kerr nonlinearity in cQED.
The role of cross-Kerr interactions has been well recognized in context of controllable quantum coherence in cQED~\cite{zhao2017two,iyama2024observation}, and have recently been used to generate tunable hopping in high-coherence quantum circuit~\cite{kounalakis2018tuneable} and high fidelity readout~\cite{wang202499,chapple2501balanced}.

Here, we use nonlinearities of a multimode high-$Q$ superconducting ring resonator. Ring resonators have been extensively investigated in optical~\cite{Kippenberg2011,del2007optical,OITKipp} and silicon platforms~\cite{razzari2010cmos}, and devices based on lithium niobate~\cite{LNOIPoberaj}.
The presence of counterpropagating modes arising from higher angular momentum states in the same resonator enable strong nonlinear interactions among the modes, which are otherwise suppressed in traditional multi-resonator systems \cite{CaoCoupledRes,DrummondandWalls}. 
We show that for nearly degenerate modes of the driven-dissipative resonator, the presence of cross-Kerr interactions can collapse the steady state into symmetric photonic configurations. 
The symmetry collapse in mode population is lifted even in the presence of a small TRSB perturbation, thus enabling TRSB sensing. 
This is particularly evident when the system is driven around parametric instabilities where even weak TRSB leads to changes in mode population configurations. As we show below, mode population asymmetry is otherwise suppressed at low and high driving amplitudes outlining the optimal regimes of device operation for high sensitivity. 
Additionally, we perform the full quantum analysis to examine the role of fluctuations near bifurcation and and quantify the signal-to-noise ratio under realistic thermal and dephasing conditions. Exploiting intrinsic nonlinearities in a multimode resonator operating in the GHz regime, we propose an on-chip and cryogenic device for probing hidden correlated order in quantum materials. 

\section{Multimode ring resonator}
We consider a superconducting ring resonator (see \autoref{fig:SystemSetup}a) that supports two degenerate counter-propagating microwave mode which arise as a consequence of odd angular momentum states, nonlinear kinetic inductance of the superconducting setup~\cite{performance_cavity_parametric},  or can be generated by bi-directional inputs to the resonator~\cite{Hill,zhang2020broadband}. The sample (green box in \autoref{fig:SystemSetup}a) is capacitively coupled to the resonator. The TRSB is encoded in the dielectric response of the sample, $E_\mu^* \epsilon_{\mu\nu}E_\nu$. The presence of antisymmetric components, $\epsilon_{\mu\nu} = -\epsilon_{\nu\mu}\neq 0 $, e.g., from Hall response would alter the transmission spectrum of the system enabling the direct probe of symmetry breaking. 
For effective manipulation of the modes, a coherent drive is applied to both modes via transmission line.
The driving process involves coupling photons from an external source into the resonator at a controlled rate, allowing the system to reach a steady state with measurable photon occupancy in each mode, as shown in \autoref{fig:SystemSetup}a.
We model the two modes of the ring resonator as two lattice sites of a Hubbard model, see \autoref{fig:SystemSetup}b, see also Supplemental Material  \cite{supplemental} for full derivation.  
The full driven-dissipative system can be described by the Hamiltonian: 
\begin{align}\label{eq:RotHam}
    \hat{H} &= \hat{H}_0 + \hat{H}_{int} + \hat{H}_{NL} + \hat{H}_{Drive} 
\end{align}
where $\hat{H}_0 = \omega_a \hat{a}^\dagger \hat{a} + \omega_b \hat{b}^\dagger \hat{b}$ is the intrinsic Hamiltonian 
, and $\hat{H}_{NL} = \frac{U}{2}(\hat{a}^{{\dagger}^{2}} \hat{a}^2 + \hat{b}^{{\dagger}^{2}} \hat{b}^2) + V(\hat{a}^\dagger \hat{a}\hat{b}^\dagger \hat{b})$ describes nonlinear interaction via the self-Kerr \textit{U} and cross-Kerr \textit{V} couplings. The operators $\hat{a}$ and $\hat{b}$ are for clockwise and anticlockwise propagating modes in the ring resonator. The two microwave modes have frequencies $\omega_a = \omega_o + \epsilon$ and $\omega_b = \omega_o - \epsilon$ where $\omega_o$ is the natural frequency of the cavity and $\epsilon$ is the natural frequency splitting between the modes.
While both self-Kerr and cross-Kerr interactions $U, V$ arise from the nonlinear kinetic inductance, the origin of V can be attributed to a change in frequency of one mode with the population of another mode which is a highly likely situation in a resonator with inherently multiple modes. In a typical cQED setup with single mode resonators, this can be achieved through Josephson junction supported coupling of two resonators~\cite{TuneableHopping}. Importantly, 
\begin{equation}
\label{eq:exchange}
    \hat{H}_{int} = (J\hat{a}\hat{b}^\dagger + J^* \hat{a}^\dagger \hat{b})
\end{equation}
is the linear exchange interaction between the modes where $\text{Im}[J]\neq 0$ signifies TRSB.
According to the coupled mode theory \cite{haus2002coupled, yariv2003coupled}, the condition for presence of TRSB follows directly from the antisymmetric dielectric response as $J\propto \int d\textbf{r} E_\mu^*(\textbf{r}) \epsilon_{\mu\nu}(\textbf{r}) E_\nu(\textbf{r})$. Finally, $\hat{H}_{Drive} = -i \sum_{j=1}^2 F_j(t) (\sqrt{\kappa_{a,j} a^\dagger} + \sqrt{\kappa_{j,b} b^\dagger}) + h.c.$ represents the drive tone applied consisting of a continous pump and probe tones. $F_j(t) = \sum_m F_{j,m} e^{-i \omega_{j,m} t}$ represents amplitudes and frequencies for both pump and probe tones. The modes are coupled to the resonator with rates $\kappa_{a}$ and $\kappa_{b}$; photon conservation assumed throughout.
Note that with $\text{Re}[J]\neq 0$ other nonlinear terms have negligible impact, previously seen in experiments in systems with linear coupling~\cite{CaoCoupledRes,rodriguez2016interaction}.

We characterize our model and highlight the role of $V$ by analyzing the energy eigenvalues of Eq.~\eqref{eq:RotHam} without a drive (see Fig.~\ref{fig:SystemSetup}b) under mean-field approximation, such that $a^\dagger a = n_a$ and $b^\dagger b=n_b$ correspond to the population of each mode. 
In Fig.~\ref{fig:SystemSetup}c, we show the degenerate $\epsilon\rightarrow 0$ case, where $V\neq 0$ enforces complete degeneracy of the modes owing to its competition with $\text{Re}[J]$. However, for  modes with $\epsilon\neq 0$, we find that the role $V$ in the system becomes less effective. This demonstrates that large modes with large frequency differences are unfavorable for the device we propose, thus, we focus on the degenerate regime. 

To understand the dynamics of our system, we define the equations of motions in the Heisenberg picture where we include dissipation coupling to the environment, which can be induced via photon loss by spontaneous decay or photon leakage:

\begin{align}\label{eq:QMaster}
    \Dot{\hat{a}} &= (-i(\omega_a + U \hat{a}^\dagger \hat{a}   + V \hat{b}^\dagger\hat{b}  + \epsilon) - \frac{\gamma}{2})\hat{a}  + iJ \hat{b} + \sqrt{\kappa} a_{in}(t) \\
    \nonumber
    \Dot{\hat{b}} &= (-i(\omega_b + U \hat{b}^\dagger\hat{b}  + V \hat{a}^\dagger \hat{a}  - \epsilon) - \frac{\gamma}{2})\hat{b}  + iJ^* \hat{a} + \sqrt{\kappa} b_{in}(t)
\end{align}
where both resonator modes $a$ and $b$ are subject to the total decay given by $\gamma = \kappa_a + \kappa_b$. 

\section{Steady state device operation and sensing TRSB}
\subsection{Symmetric Collapse and Steady State}
\begin{figure}
    \centering
    \includegraphics[width = \linewidth]{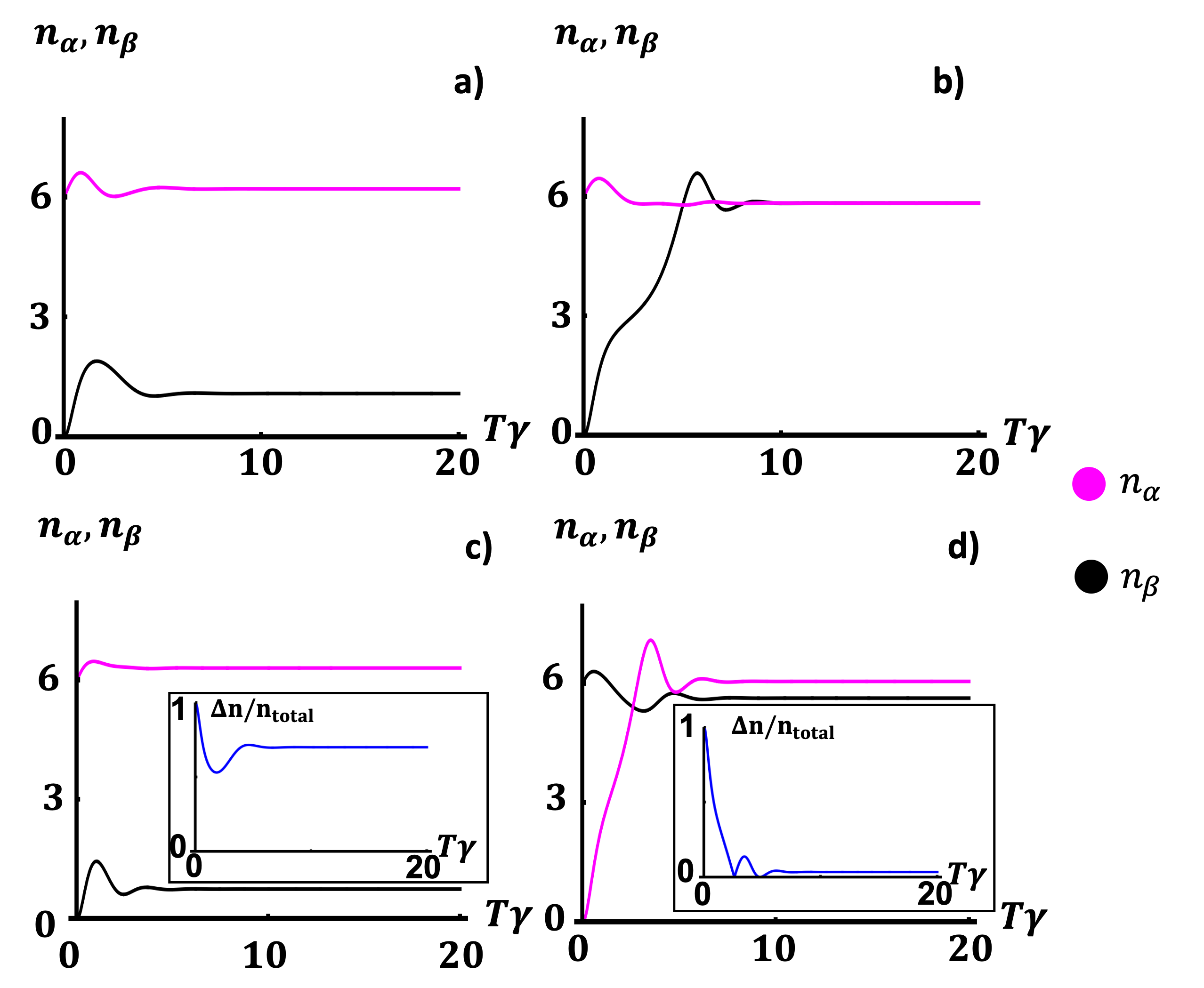}
    \caption{\justifying Long time dynamics of the driven-dissipative, two mode resonator obtained from semi-classically solving Eq.~(\ref{eq:QMaster}). 
    (a) Semiclassical steady state mode populations for $V=0$ \addAA{and $V=0.1$; $n_\alpha =6.0$, and $n_\beta = 0.0$.} 
    (c) The symmetry collapse in populations is strongly lifted when TRSB is introduced into the model, i.e.,  $\text{Im}[J] \neq 0$ ($\text{Im}[J] = 0.1$)
    (d) Nonreciprocal behavior of splitting mode populations with respect to initial mode populations, $n_\alpha = 0.0$ and $n_\beta = 6.0$. Other parameter used: $ U=0.6$, $\text{Re}[J] = 0.1$, $\kappa = 1$, $\Delta = -3.5$ and $\gamma = 2$. All parameter values are given in units of dissipation $\gamma$.}
    \label{fig:steady-state}
\end{figure}

To demonstrate the resonators ability due to $V\neq 0$, we solve for the steady state of the system.
In Fig.~\ref{fig:steady-state}a,b we show the long time dynamics of $n_{\alpha,\beta}$ for a fixed drive amplitude with $V=0$ and $V\neq 0$, respectively. 
Strikingly, the presence of a cross-Kerr nonlinearity $V$ enables a symmetric collapse of photon populations in steady state even when they are subjected to  different initial conditions. 
This overriding of asymmetric initial conditions into a final symmetric state highlights the role of nonlinear cross-mode interactions in enhancing mode correlations.
This enables a sensitive probe for TRSB as we note the transition to asymmetric photon populations even for small values of $\text{Im}[J]$, see Fig. ~\ref{fig:steady-state}c. 
Interestingly, the splitting of mode population configurations upon TRSB is highly nonreciprocal with respect to the initial mode population, a feature unique to our proposed device, as shown in Fig. ~\ref{fig:steady-state}d where the initial mode populations are reversed when compared to Fig.~\ref{fig:steady-state}c. We qualitatively quantify the nonreciprocal sensitivity to TRSB by plotting the ratio $\Delta n/n_{\rm total}$ where $\Delta n = n_\alpha-n_\beta$ and $n_{\rm total} = n_\alpha + n_\beta$. We note $\Delta n/n_{\rm total} \approx 70\%$ versus 10\% upon switching of the initial conditions, see inset of Fig. ~\ref{fig:steady-state}c,d.

\begin{figure*}
    \centering
    \includegraphics[width =\linewidth]{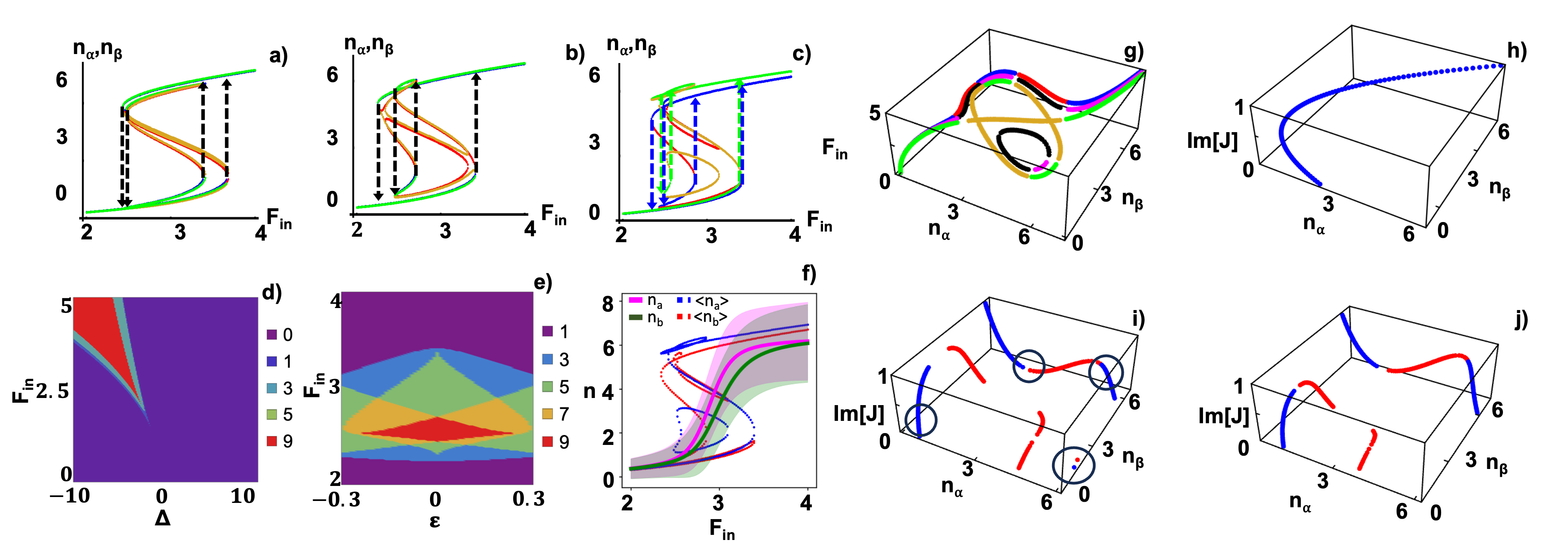}
    \caption{\justifying Parametric instabilities and TRSB sensitivity beyond the bifurcation threshold. (a,b) Multistable steady state solutions for populations of each mode when (a) $V=0$, (b) $V=0.1$. Here, blue (green) denote stable solutions for mode population $n_\alpha (n_\beta)$ and red (gold)  denote unstable solutions for mode population $n_\alpha (n_\beta)$.  
    (c) For $\text{Im}[J] = 0.1$, new configurations are obtained, where the two modes undergo \addAA{low-high} transitions at different drives, blue and green dashed arrows.
    (d-e) Optimal parameters by determining the number of steady-state solutions in $\Delta/F_{in}$ (d) and $\epsilon/F_{in}$ (e) phase space, \addAA{indicating senstive device operator for red detuning $\Delta < 0$, and nearly degenerate modes, $\epsilon \rightarrow 0$.}
    (f) \addAA{Comparison of Semiclassical (red, blue dots) and full quantum (pink, dark-green curves) solutions of photonic population in variation with $F_{in}$. Shaded regions indicate the instantaneous quantum variance $\sigma_{shot}$}.
    (g) 3D visualization of $n_\alpha$ and $n_\beta$ on $F_{in}$. The stable (unstable) solutions are shown in green (gold) for $\text{Im}[J] = 0$, magenta (black) for $\text{Im}[J]=0.1$, and blue (red) for $\text{Im}[J]=0.2$. When $\text{Im}[J]\neq 0$, the number of stable solutions increases for asymmetric mode configurations. (h-i) Behavior of photon occupation in variation with $\text{Im}[J]$ for $F_{in}=1.8$, $F_{in}=2.5$ and $F_{in}=3.0$, respectively. 
    Other parameter values are the same as Fig.~\ref{fig:steady-state}.}
    \label{fig:classical}
\end{figure*}

\subsection{Operating Regime and Stability Analysis}
It is useful to operate the device in a regime near the bifurcation threshold where small perturbations can lead to drastic shifts in the steady state photon occupation numbers. We analyze the variation of steady state photon populations under variations of drive amplitude $F_{in}$. 
The evolution of parametric instabilities with drive amplitude is shown in Fig.~\ref{fig:classical}(a,b). Stability is determined by linearizing the equations of motion around the mean-field solutions and analyzing the eigenvalues of the resulting stability matrix, see Supplemental Material \cite{supplemental}. 

\addND{For $V=0$, the system exhibits the familiar bistable branches of coupled nonlinear resonators, corresponding to low- and high-occupation configurations of the two modes~\cite{CaoCoupledRes}. When $V\neq0$, these branches hybridize due to the symmetric collapse mechanism described above, producing a modified multistable landscape that is highly sensitive to TRSB.} 

In Fig.~\ref{fig:classical}c, we show $n_\alpha$ and $n_\beta$ as a function of $F$ for $\text{Im}[J]>0$ and find new emergent mode population configurations which demonstrate signatures of TRSB, reminiscent of the system transitioning from symmetric to asymmetric configurations in Fig.~\ref{fig:steady-state}c. 
For $\mathrm{Im}[J]\neq0$, new asymmetric steady-state branches emerge that are absent in the symmetric case, and the population splitting becomes more pronounced at higher photon numbers, providing a clear operating window for sensing.

Importantly, the difference in mode populations is small when $F_{in}$ is very small or very large, indicating that the region of multistable solutions is the optimal parameter space for device operation. Additionally, here we use the same $U$ for both modes; steady-state analysis for modes having different $U$ is shown in Supplemental Material \cite{supplemental}. To identify optimal operating conditions, we map the number of steady-state solutions in the $\Delta/F_{in}$ and $\epsilon/F_{in}$ parameter spaces (Fig.~\ref{fig:classical}d,e) -- multistability is maximized for red detuning and nearly degenerate modes.

Next, we visualize in 3D the photon population of each mode. In Fig.~\ref{fig:classical}g, $n_\alpha$ and $n_\beta$ are shown as a function of $F_{in}$. We show stable (unstable) points by green (gold), magenta (black) and blue (red) for $\text{Im}[J]=0$, $\text{Im}[J] =0.1$ and $\text{Im}[J]=0.2$, respectively.  Similar to Fig.~\ref{fig:classical}b-c, we find drastic changes around the region of parametric instabilities which can be exploited to probe TRSB. Importantly, the 3D visualizations help distinguish the spontaneously symmetry broken states (multistable solutions for $\text{Im}[J]=0$) from TRSB configurations. 
Furthermore, any instances of spontaneous symmetry breaking can be averaged out through multiple iterations of measuring the photon population, thus allowing us to distinguish between TRSB and any other form of spontaneous nonlocal interaction. The dependence of photon populations on $\mathrm{Im}[J]$ is shown in Fig.~\ref{fig:classical}h–j for different drive amplitudes. For small drive ($F_{in}=1.8$), Fig~\ref{fig:classical}h, only a single symmetric steady state is present and no measurable TRSB response emerges. Near the bifurcation regime ($F_{in}=2.5$),Fig~\ref{fig:classical}i, multiple steady-state branches appear, and even weak TRSB perturbations induce transitions to asymmetric configurations, providing a sensitive operating point. 
For larger drive powers ($F_{in}=3$), Fig.~\ref{fig:classical}j, the number of accessible configurations decreases and asymmetric states appear only for large $\mathrm{Im}[J]$, reducing the sensitivity of the device. These results confirm that optimized driven–dissipative operation near the bifurcation threshold provides the highest sensitivity to weak TRSB signals in the sample.

\section{Quantum Analysis}

\begin{figure*}
    \centering
    \includegraphics[width=\linewidth]{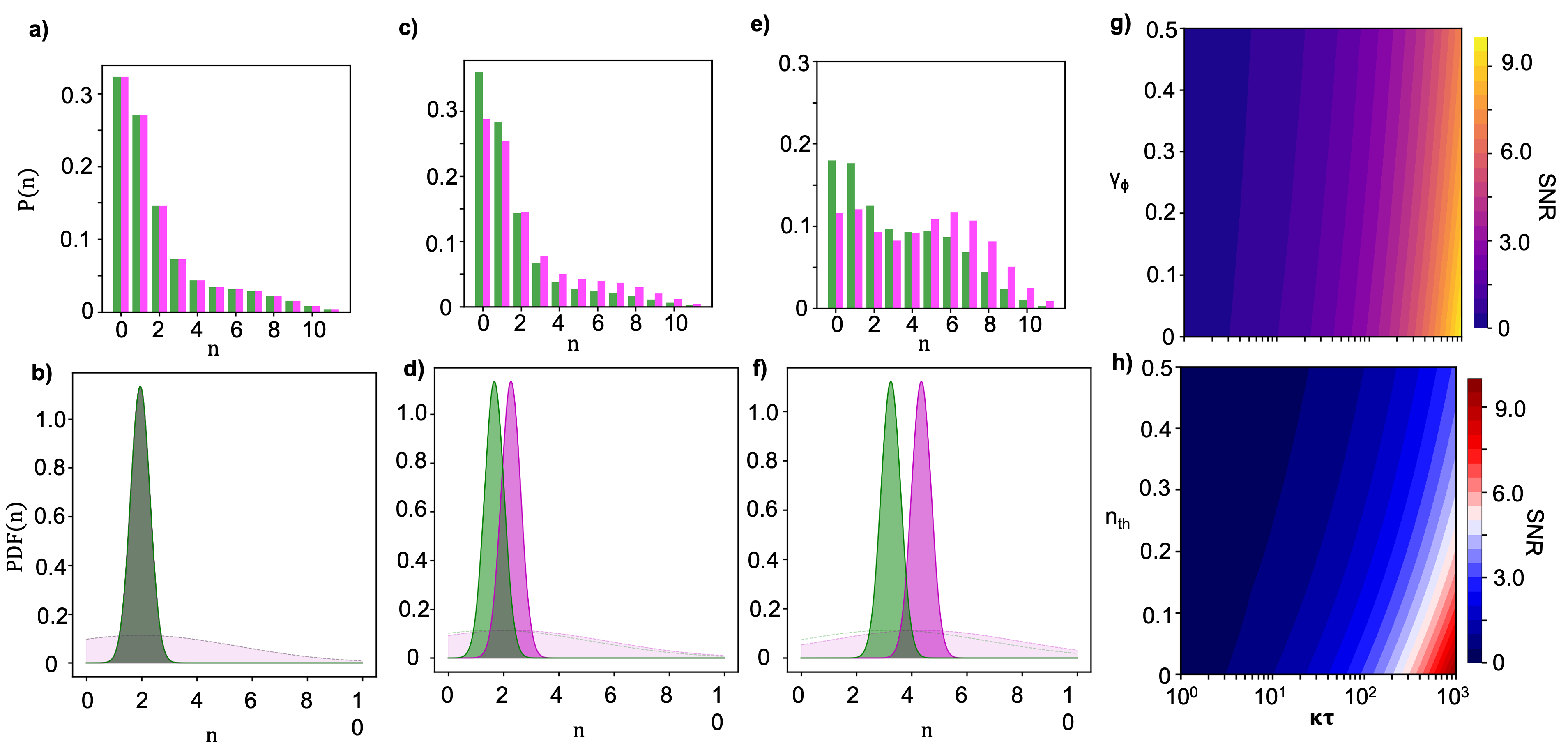}
    \caption{\justifying (a,c,e) Photon number probability, $P(n)$  and (b,d,f) probability density function (PDF) obtained from quantum analysis for Mode A (green) and Mode B (magenta) under varying interaction strengths ($V$) and TRSB signals ($\text{Im}[J]$); faint dashed lines represent instantaneous single-shot distributions (broadened by quantum fluctuations), while solid lines show the time-integrated distributions ($\kappa\tau=50$). (a,b) $\text{Im}[J]=0, V=0$, (c,d) $P(n)$ and PDF for $\text{Im}[J]=0.1, V=0$ and (e,f) $\text{Im}[J]=0.1, V=0$; $V\neq 0 $ resolves the photonic distribution up-converting the magnetic effects in the material-resonator hybrid system. (g-h) Robustness to Experimental Noise. Contour maps of the integrated SNR as a function of measurement integration time $\tau$ and experimental noise sources.
    (g) Dephasing: Pure dephasing ($\gamma_{\phi}$) reduces the effective cooperativity and signal contrast. The device demonstrates resilience, maintaining $\text{SNR} > 5$ for dephasing rates up to $\gamma_{\phi} \approx 0.1\kappa$ for few shot measurements, validating the feasibility of the scheme using current superconducting circuit technology.
    (h) Thermal Noise: Increasing the bath temperature ($n_{th}$) broadens the Wigner distribution, requiring longer integration times to resolve the signal. However, even with significant heating ($n_{th} \approx 0.2$), high-fidelity detection is achievable within standard readout windows ($\kappa \tau \sim 10^2$).}
    \label{fig:quantum}
\end{figure*}

\subsection{Fluctuation Analysis and Resolution}
While the semiclassical analysis provides a transparent understanding of TRSB sensing using ring resonators,
prior studies of driven-dissipative dimers \cite{CaoCoupledRes} have shown that fluctuations diverge near critical points, potentially destabilizing the steady states. To validate the sensing capability of the ring resonator in the quantum limit, we solve the full Lindblad Master Equation~\cite{quantum_sensing}:
\begin{equation}
    \Dot{\rho} = -\frac{i}{\hbar} [H_{eff}, \rho] + \sum_i \gamma_i (L_i \rho L_i^\dagger - \frac{1}{2} \{L_i^\dagger L_i,\rho\}) 
\end{equation}
where the index $i$ sums over the two counterpropagating modes in the ring resonator. 

In Fig.~\ref{fig:classical}f, we show the full quantum solutions of mode populations (pink and drak-green lines) along with semiclassical solutions (red and blue dots) as obtained in Fig.~\ref{fig:classical}c. The shaded region shows the corresponding quantum variance, $\sigma_{shot} = \sqrt{{\rm Var}(\Delta n)}$~\cite{quantum_sensing}. While we note that fluctuations washout the multistable states, the conclusion that maximum separation in mode populations for $\text{Im}[J]\neq 0$ occurs beyond bifurcation regime remains robust even though fluctuations are maximum around the bifurcation regime.

We further analyze the $V$ enabled advantage in ring resonators by simulating multi-shot measurements which form the basis of schemes based on time-integrated detection.
The noise floor for such a measurement is governed by the Standard Error of the Mean (SEM), which scales inversely with the integration time $\tau$ such that 
$\sigma_{m}(\tau) \approx \sigma_{shot}/\sqrt{\kappa \tau}$~\cite{quantum_sensing}. 

In Fig.~\ref{fig:quantum}, we analyze the probability of photon occupation $P_i(n) = {\rm diag}[{\rm Tr}_i\rho]$ and the corresponding probability density function (PDF), see Supplemental Material \cite{supplemental} for details on the definition of PDF used in this work. 
For $\text{Im}[J]=0$, see Fig.~\ref{fig:quantum}a,b the system resides in a symmetric configuration. Interestingly, we find stark split in $P_i(n)$ and PDF for $V=0.1$ (panels e,f) compared to $V=0$ (panels c,d).  
This behavior reflects the emergence of interaction-stabilized metastable manifolds: the nonlinear term $Vn_a n_b$ energetically penalizes simultaneous occupation of both modes, effectively sharpening the potential landscape and suppressing switching between symmetric and asymmetric configurations.

Cross-Kerr initiates a nonlinear feedback mechanism that suppresses fluctuations and improves semiclassical agreement, see Supplemental Material \cite{supplemental} for comparison of semiclassical and quantum analysis with corresponding quantum fluctuations. In a setup with multi-shot measurements, $V\neq 0$ up-converts the magnetic effects in our ring resonator device improving the distinguishability of the photon populations, underpinning the sensing mechanism proposed here.

\subsection{Experimental Benchmarks}
To assess the robustness of this sensing mechanism, we model the effects of noise within the Lindblad master-equation framework by including photon loss, thermal excitation, and pure dephasing processes. Photon decay at rate $\kappa$ and thermal occupation $n_{th}$ broaden the photon-number distributions and increase the variance of the population imbalance, while dephasing $\gamma_{\phi}$ introduces additional diffusion in phase space. Typical values of thermal occupations are $n_{th} \leq 10^{-3} - 10^{-2}$ and pure-dephasing rates around $\gamma_{\phi} \leq 0.01 -0.1\kappa$ ~\cite{lu2023resolving}. 
Fig. ~\ref{fig:quantum}g,h maps the integrated SNR versus thermal occupation $n_{th}$ and dephasing rate $\gamma_\phi$ as a function of integration time. In both cases the SNR increases with averaging time, while dephasing and thermal occupation primarily reduce the achievable SNR by broadening the photon-number distributions, with thermal noise producing the stronger degradation at large noise levels.

\addND{The parameter regimes considered here are consistent with experimentally realized superconducting microwave resonators and multimode circuit-QED devices. Typical linewidths lie in the range $\kappa/2\pi \sim 0.1-5$ MHz, corresponding to quality factors $Q \sim 10^4 - 10^6$ depending on geometry and materials. Kerr nonlinearities arising from Josephson elements or engineered nonlinear resonators are routinely observed in the range $U/2\pi \sim 10-100$ kHz, with comparable or larger effective cross-Kerr couplings achievable in multimode or strongly coupled architectures. Experiments have demonstrated photon-number–dependent shifts and cross-Kerr interactions between microwave modes at the single-photon level, confirming that interaction strengths in this range are realistic for superconducting devices ~\cite{kirchmair2013observation, hillmann2022designing, holland2015single}. Coupling rates between resonator modes or hybridized excitations typically fall in the range $J/2\pi \sim 0.1-10$ MHz, while effective nonreciprocal or symmetry-breaking perturbations corresponding to $\text{Im[J]}$ are generally expected to be small compared to the real coupling, often at the level of a few percent or less of J, consistent with perturbative treatments of nonreciprocal or gyrotropic interactions in microwave photonic systems. Drive amplitudes corresponding to intracavity photon numbers in the range $n\sim 1-10$ are routinely accessible in circuit-QED experiments without inducing unwanted heating or quasiparticle generation, and nonlinear instabilities and bifurcation phenomena have been observed in this same photon-number regime in Josephson-based resonators ~\cite{bera2024single,bosman2017approaching,baust2016ultrastrong}.}  \\

\addND{Within experimental values of these orders such as detunings of a few linewidths $\Delta/\kappa \sim 1-10$, near-degenerate mode splittings $\epsilon<<\kappa$, our simulations indicate that the interaction-enhanced population imbalance remains resolvable over integration times $\kappa \tau \sim 10^1-10^3$. These values correspond to physical integration times of order $0.1-10 \mu s$ for typical microwave resonators, well within standard dispersive readout capabilities. Taken together, these estimates demonstrate that the proposed multimode superconducting ring resonator operates in a parameter regime that is not only experimentally accessible but also naturally suited to exploiting cross-Kerr–stabilized metastability as a sensitive probe of weak TRSB signatures in quantum materials.}

\section{Conclusion}
\label{section:Conclusion}

In this work, we explore the potential of a multimode superconducting ring resonator within the framework of cQED for detecting TRSB signatures in quantum materials. 
By leveraging the interplay of nonlinear cross-Kerr interactions, self-Kerr nonlinearities, and driven-dissipative dynamics, we demonstrate the sensitivity of the system to TRSB-induced perturbations through the imaginary component of the exchange coupling parameter, $\text{Im}[J]$, which results in transitions of the system from symmetric to asymmetric photon configuration states. These symmetry-breaking transitions, enabled by the unique nonlinear dynamics of the system, provide a reliable theoretical signature, subject to further experimental validation.\\

The cross-Kerr term $V$ in a single cavity, which plays a central role in this work, is shown to induce parametric instabilities and multistability regimes, even in the absence of traditional coupled cavity systems. These effects expand the operational versatility of our device and allow for precise control over mode population dynamics. 
The observed transitions in mode populations underline the practical applicability of the device for experimental detection of TRSB in real-world systems.  Our findings establish that the proposed multimode resonator operates effectively in GHz regimes for resonator and drive frequencies, with nonlinearities in the kHz range \cite{level_attraction_idler_resonance} and interaction strengths in the MHz range ~\cite{magnon, OITKipp}. These operating regimes ensure compatibility with current experimental paradigms and emphasize the practicality of the approach.

To validate the sensing mechanism beyond the semiclassical approximation, we further carried out a full quantum analysis based on the Lindblad master equation. While quantum fluctuations smooth out idealized bistability in the long-time limit, we find that the TRSB-induced population imbalance remains finite and experimentally resolvable. The cross-Kerr interaction plays a crucial role in this regime by stabilizing metastable configurations, reducing effective noise, and enhancing the distinguishability of photon populations. These results demonstrate that the proposed sensing mechanism remains robust in the quantum regime and does not rely on idealized mean-field behavior.

Unlike traditional Kerr rotation which probes TRSB via polarization rotation of far fields, our approach relies on leveraging inter-mode coupling and intrinsic nonlinear interactions in superconducting devices. Additionally, given that magneto-optic effects are suppressed at microwave frequencies, our proposed device emerges as a powerful TRSB tool in the microwave domain. Though recent proposals for microwave Kerr rotations using Sagnac interferometry have been put forward \cite{chouinard2025probing}, their on-chip compatibility is questionable. On the contrary, our proposed device possesses no such limitations. 

More broadly, this work highlights the potential of superconducting resonators not only as elements of quantum information processors but also as precision diagnostic tools for probing exotic phases of matter. By focusing on a single multimode cavity, we extend existing approaches that predominantly rely on coupled resonator arrays and self-Kerr nonlinearities \cite{quantum_limited_amplification, level_attraction_idler_resonance}. Future research can build on this foundation by integrating such devices into networks of resonators \cite{mcbroom2024entangling,miao2024kerr} and exploring coupling to a wider range of quantum materials. These directions may ultimately enable superconducting resonators to become versatile platforms for precision sensing and quantum-material diagnostics.\\

\begin{acknowledgments}
We acknowledge helpful discussions with Saulius Vaiteikenas, Kerry (Kandgi) Yu, Roberto Negrin, Jugal Talukdar, Aman Mehta, William Munizzi, and Jack Diab. This work was supported by the Quantum Science Center (QSC), a National Quantum Information Science Center of the U.S. Department of Energy (DOE). We also acknowledge Grant Numbers GBMF8048 and GBMF12976 from the Gordon and Betty Moore Foundation.
\end{acknowledgments}

\vspace{0.1cm}
\section*{Data and Code Availability}
The data and code that support the findings of this study are available from the corresponding authors upon reasonable request. 

\bibliography{references}

\appendix
\onecolumngrid
\newpage
\pagebreak
\widetext
\setcounter{equation}{0}
\setcounter{figure}{0}
\setcounter{table}{0}
\setcounter{page}{1}
\makeatletter
\renewcommand{\theequation}{S\arabic{equation}}
\renewcommand{\thefigure}{S\arabic{figure}}
\renewcommand{\bibnumfmt}[1]{[S#1]}

\section*{Supplemental Material for ``Probing Time Reversal Symmetry Breaking using a Nonlinear Superconducting Ring Resonator''}

\section{Detailed Derivation of Hamiltonian} \label{appendix-A}
In our case, the two microwave modes are considered to be quasi-degenerate such that their intrinsic Hamiltonian is $(\hbar = 1)$:

\begin{equation}\label{eq:Ham0}
    H_0 = \omega_a \hat{a}^{\dagger} \hat{a} + \omega_b \hat{b}^{\dagger} \hat{b}
\end{equation}

where $\omega_a = \omega_0 + \epsilon$ and $\omega_b = \omega_0 - \epsilon$ are the resonant cavity frequencies with $\omega_o/2\pi = 4.95$ GHz.  Furthermore, $\hat{a} (\hat{b})$ and $\hat{a}^{\dagger} (\hat{b}^{\dagger})$ are the annihilation and creation operators of mode a (mode b), respectively. The ring resonator, being superconducting, has an associated nonlinearity coming from its kinetic inductance. This kinetic inductance of our system will lead to nonlinear effects in our system such that a self- (U) and cross-Kerr (V) term will appear. The nonlinear Hamiltonian part coming from the kinetic inductance is:

\begin{equation}\label{eq:NLHam}
    H_{NL} = \frac{U}{2} (\hat{a}^{{\dagger}^{2}} \hat{a}^2 + \hat{b}^{{\dagger}^{2}} \hat{b}^2) + V (\hat{a}^{\dagger} \hat{a}  \hat{b}^{\dagger} \hat{b})
\end{equation}

These terms are equivalently to be understood as on-site interaction strength and inter-site interaction strength in a Bose-Hubbard system. Furthermore, we consider linear interactions in the system as a hopping parameter between the two sites. Only terms of equal number of creation and annihilation operators are kept to ensure equal exchange between both sites, similar to the hopping rate in a Bose-Hubbard dimer.  We label the coupling strength (hopping rate) between both modes as $J$:
\begin{equation}\label{eq:RWAHam}
    H_{int} = (J\hat{a} \hat{b}^\dagger +J^* \hat{a}^{\dagger} \hat{b})
\end{equation}

In order to control, examine the state and for the modes to have non-zero value in the steady-state, of the microwave resonator, we need to look at external couplings or dissipation since we so far have only considered interaction between the modes. We couple the microwave resonator to waveguides allowing for photons to enter and escape the cavity. An external drive is applied to the microwave resonator. In our case, a monotonic coherent drive tone is applied to both resonator modes that induces the drive Hamiltonian:

\begin{equation}\label{eq:Hdrive}
    \hat{H}_{drive} = -i \sqrt{\kappa} (F_{in} (t) (\hat{a}^{\dagger} + \hat{b}^{\dagger}) e^{-i\omega_d t} + h.c)
\end{equation}

where $F_{in}(t)$ represents the coherent tone applied to the resonator via a transmission line at drive frequency $\omega_d$ as well as the probe tones applied. $\kappa$ represents the external coupling rate from our transmission line to our resonator, where the exchange in photons between the resonator and transmission line will add to the total decay rate of our system. We can now go into the rotating frame by applying the unitary evolution $\hat{H}_{rot} = \hat{U} \hat{H} \hat{U}^{\dagger} -i\hat{U}\Dot{\hat{U}}^{\dagger}$, where $\hat{U} = e^{i \omega_d (\hat{a}^{\dagger} \hat{a} + \hat{b}^{\dagger} \hat{b})t}$. In the rotating frame, our Hamiltonian becomes 

\begin{align}\label{eq:RotHam2}
    \hat{H}_{rot} &= \Delta (\hat{a}^{\dagger} \hat{a} + \hat{b}^{\dagger} \hat{b}) + \epsilon (\hat{a}^{\dagger} \hat{a} - \hat{b}^{\dagger} \hat{b})\\ 
    &+ U \left((\hat{a}^{\dagger})^2 \hat{a}^2 + (\hat{b}^{\dagger})^2 \hat{b}^2\right) \nonumber + V (\hat{a}^{\dagger} \hat{a}  \hat{b}^{\dagger} \hat{b})\\
    & + J(\hat{a}^{\dagger} \hat{b} + \hat{a} \hat{b}^{\dagger}) - i \sqrt{\kappa} (F_{in} (t) (\hat{a}^{\dagger} + \hat{b}^{\dagger}) + h.c.)
\end{align}

Here $\Delta =\omega_0 -\omega_d$ represents the detuning of our cavity with the resonant drive. 

\section{Analytical derivation of semiclassical state equations}
\label{appendix-B}
When analyzing the steady state solution of equations (\ref{eq:QMaster}), we write the semiclassical equations for the complex modes $\alpha$ and $\beta$ and their respective conjugates:
\begin{align}
    \Dot{\alpha} &= (-i(\Delta + U n_\alpha + V n_\beta  + \epsilon) - \frac{\gamma}{2})\alpha  + iJ \beta + \sqrt{\kappa} F_{in} \\
    \Dot{\alpha^*} &= (i(\Delta + U n_\alpha + V n_\beta  + \epsilon) - \frac{\gamma}{2})\alpha^*  - iJ^* \beta^* + \sqrt{\kappa} F_{in}^* \\
    \Dot{\beta} &= (-i(\Delta + U n_\beta  + V n_\alpha  - \epsilon) - \frac{\gamma}{2})\beta  + iJ \alpha + \sqrt{\kappa} F_{in} \\
    \Dot{\beta^*} &= (i(\Delta + U n_\beta  + V n_\alpha  - \epsilon) - \frac{\gamma}{2})\beta^*  - iJ^* \alpha^* + \sqrt{\kappa} F_{in}^* 
\end{align}
where $n_\alpha = |\alpha|^2$ and $n_\beta = |\beta|^2$ are the photon occupation numbers for each mode. 
We now set all time derivatives to zero $\frac{d \alpha}{d t} = \frac{d \beta}{d t} = 0$ and solve for $\alpha_s$ and $\beta_s$:

\begin{align}\label{eq:SSEOM}
    0 &= \big(-i(\Delta + Un_\alpha + Vn_\beta + \epsilon) - \frac{\gamma}{2}\big)\alpha_s 
     + iJ\beta_s + \sqrt{\kappa} F_{in}, \\
    0 &= \big(i(\Delta + U n_\alpha + Vn_\beta + \epsilon) - \frac{\gamma}{2}\big)\alpha_s^{*}
     - iJ^*\beta_s^{*} + \sqrt{\kappa} F_{in}^{*}, \\
    0 &= \big(-i(\Delta + Un_\beta + Vn_\alpha - \epsilon) - \frac{\gamma}{2}\big)\beta_s 
     + iJ\alpha_s + \sqrt{\kappa} F_{in}, \\
    0 &= \big(i(\Delta + Un_\beta + Vn_\alpha - \epsilon) - \frac{\gamma}{2}\big)\beta_s^{*} 
     - iJ^*\alpha_s^{*} + \sqrt{\kappa} F_{in}^{*}.
\end{align}

We now multiply equations B5 and B6 together as well as B7 and B8. This results in two state equations of the form:

\begin{align}\label{eq:twostate}
    \kappa |\Bar{F}_{in}|^2 &= |\mu|^2 n_\alpha + |J|^2 n_\beta \\ \nonumber
    & - iJ^* \beta_s^{*} (-i(\Delta + Un_\alpha + Vn_\beta + \epsilon) - \frac{\gamma}{2})\alpha_s \\ \nonumber
    & + iJ \beta_s (i(\Delta + U n_\alpha + V\beta + \epsilon) - \frac{\gamma}{2})\alpha_s^{*}   \\
     \kappa |\Bar{F}_{in}|^2 &= |\chi|^2 n_\beta + |J|^2 n_\alpha - iJ \chi^{*} \\ \nonumber
     & - iJ^*\alpha_s^{*} (-i(\Delta + Un_\beta + Vn_\alpha - \epsilon) - \frac{\gamma}{2})\beta_s \\ \nonumber
     & i J \alpha_s (i(\Delta + Un_\beta + Vn_\alpha - \epsilon) - \frac{\gamma}{2})\beta_s^{*}
\end{align}

where $|\mu|^2 = |\Delta + Un_\alpha +Vn_\beta + \epsilon|^2  + \frac{\gamma^2}{4}$, $|\chi|^2 = |\Delta + Un_\beta +Vn_\alpha - \epsilon|^2  + \frac{\gamma^2}{4}$. Equations B9 and B10 allow us to determine the multistability regimes in terms of applied drive power $F_{in}$ or other parameters of interest.

\section{Microscopic Dynamic Analysis}\label{appendixC}

The linearized equations of motion for the fluctuations are given as

\begin{align}\label{eq:linearized}
    \Dot{\delta \alpha} &= \left( -i(\Delta + \epsilon + 2U|\alpha_s|^2 + V|\beta_s|^2) - \frac{\gamma}{2} \right) \delta \alpha 
    - i U \alpha_s^2 \delta \alpha^{\dagger} \nonumber \\
    &\quad - i(V\alpha_s \beta_s^{*} + J) \delta \beta 
    - i V \alpha_s \beta_s \delta \beta^{\dagger} 
    + \sqrt{\kappa} F_{in}  \\
    \nonumber
    \Dot{\delta \beta} &= \left( -i(\Delta - \epsilon + 2U|\beta_s|^2 + V|\alpha_s|^2) - \frac{\gamma}{2} \right) \delta \beta 
    - iU \beta_s^2 \delta \beta^{\dagger} \nonumber \\
    &\quad - i(V\alpha_s^{*} \beta_s + J) \delta \alpha 
    - i V \alpha_s \beta_s  \delta \alpha^{\dagger} 
    + \sqrt{\kappa} F_{in}  
\end{align}

Equations C1 and C2 show the linearized equations around fixed points, where the analysis of their stability becomes easier and allows us to determine how the photon occupation number in a given state deviates significantly from expected values. 
 We now solve for the stability of the fixed points using the linearized equations of motion around the fluctuations. The drift matrix is defined by the following equation relation $\Dot{u(t)} = A u(t)$, where $u(t)$ represents the state vector. The drift matrix resulting from this definition is the following:

{\footnotesize \begin{equation}\label{eq: Jacobian}
    A = \begin{pmatrix}
        -i(\Tilde{\Delta}_a  - i\frac{\gamma_a}{2}) & -i(J+V\alpha \beta^*) & ig_a & ig_x\\
        -i (J+V\alpha^* \beta) & -i (\Tilde{\Delta_b} - i\frac{\gamma_b}{2}) & ig_x & ig_b \\
        -ig_a^* & -ig_x^* & i(\Tilde{\Delta_a} + i\frac{\gamma_a}{2}) & i(J+V\alpha \beta^*)^* \\
        -ig_x^* & -ig_b^* & i(J+V\alpha^* \beta)^* & i (\Tilde{\Delta_b} - i\frac{\gamma_b}{2})
    \end{pmatrix}
\end{equation}}

where $\Tilde{\Delta_a} = \Delta_a + \epsilon + 2|\alpha|^2 U + V|\beta|^2$, $\Tilde{\Delta_b} = \Delta_b - \epsilon + 2|\beta|^2 U + V|\alpha|^2$, $g_a = U \alpha^2$, $g_b = U \beta^2$ and $g_x = V \alpha \beta$. 
When the real part of the eigenvalues of A is positive, the solution is stable. Furthermore, we look at the determinant and trace, for which the Hurwitz criterion for stability requires that the trace and determinant of this matrix be non-zero and positive for stable eigenvalues. 

\section{Additional plots on steady state conditions and 2D visualization of mode populations} \label{additional plots semiclassical}

When exploring the parameter space, the self-Kerr U and cross-Kerr term V need to be carefully tuned to be able to sense signatures of TRSB. When tuning the cross-Kerr term V to obtain collapse of the photon population to a symmetric state regardless of the initial conditions, we sweep over a couple of values of V to demonstrate its effect.

\begin{figure}[h]
    \centering
    \includegraphics[width=0.75\linewidth]{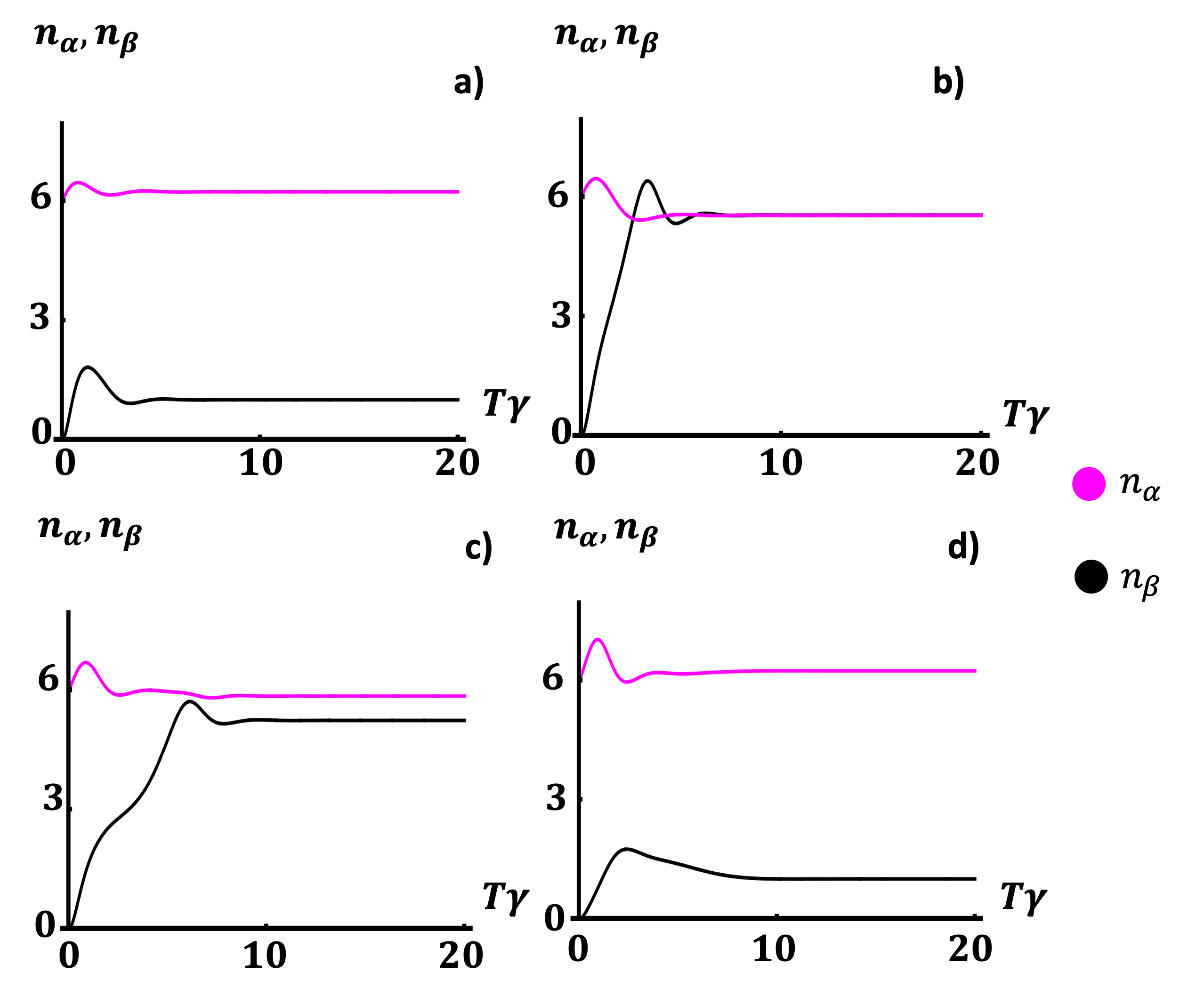}
    \caption{\justifying Long time dynamics driven-dissipative two model resonator obtained from semi-classically solving Eq.~(\ref{eq:QMaster}) with different vales of cross Kerr term V. (a) Semiclassical steady state mode population for $V=0.05$. (b) For $V = 0.15$ the mode population collapse into a symmetric configuration faster than for ~\autoref{fig:steady-state} (b). (c) The symmetry in population is slightly lifted when TRSB is introduced to the model, i.e., $\text{Im}[J] \neq 0$ (here we use $\textbf{Im}[J] = 0.15$) for $V=0.15$ indicating the requirement for careful tuning of V. (d) For stronger cross Kerr term $V=0.25$, even larger effects of $\text{Im}[J] = 0.5$ required to strong lift of mode splitting. Other parameters used: $ V=0.1$, $\text{Re}[J] = 0.1$, $\kappa = 1$, $\Delta = -3.5$ and $\gamma = 2$. All parameters are given in units of dissipation $\gamma$. }
    \label{fig:diff V}
\end{figure}

In ~\autoref{fig:diff V} we plot four different instances of V. In ~\autoref{fig:diff V} (a) we tune V to be $V=0.05$ showing that no photon collapse occurs for highly asymmetric initial conditions. Once tuning V to $V=0.15$ as in ~\autoref{fig:diff V}(b) the collapse becomes quickly and occurs faster than for $V=0.1$ as seen in ~\autoref{fig:steady-state}(b). Furthermore for stronger $V=0.15$ when having $Im[J] \neq 0$ as in ~\autoref{fig:diff V} (c) the system will not split back into the asymmetric configuration but a small splitting occurs. Thus, stronger signatures of TRSB are required to see significant effects on the mode population of the resonator. This can be seen with ~\autoref{fig:diff V} (d) where we chose an even larger $V=0.25$ and required $Im[J] = 0.5$ for the modes to be split into an asymmetric configuration again due to TRSB. Effectively the cross-Kerm term V needs to be placed in a sweet spot in order to efficiently probe materials for TRSB signatures. 

The system can be also probed for different values of drive amplitude and self-Kerr terms applied to the individual modes, i.e., we set $U_1 \neq U_2$ and $F_1 \neq F_2$. We probe the system for different conditions of self-Kerr term where the values are not too far apart and for drive terms that drive the modes at different amplitudes. 

\begin{figure}[h]
    \centering
    \includegraphics[width=0.75\linewidth]{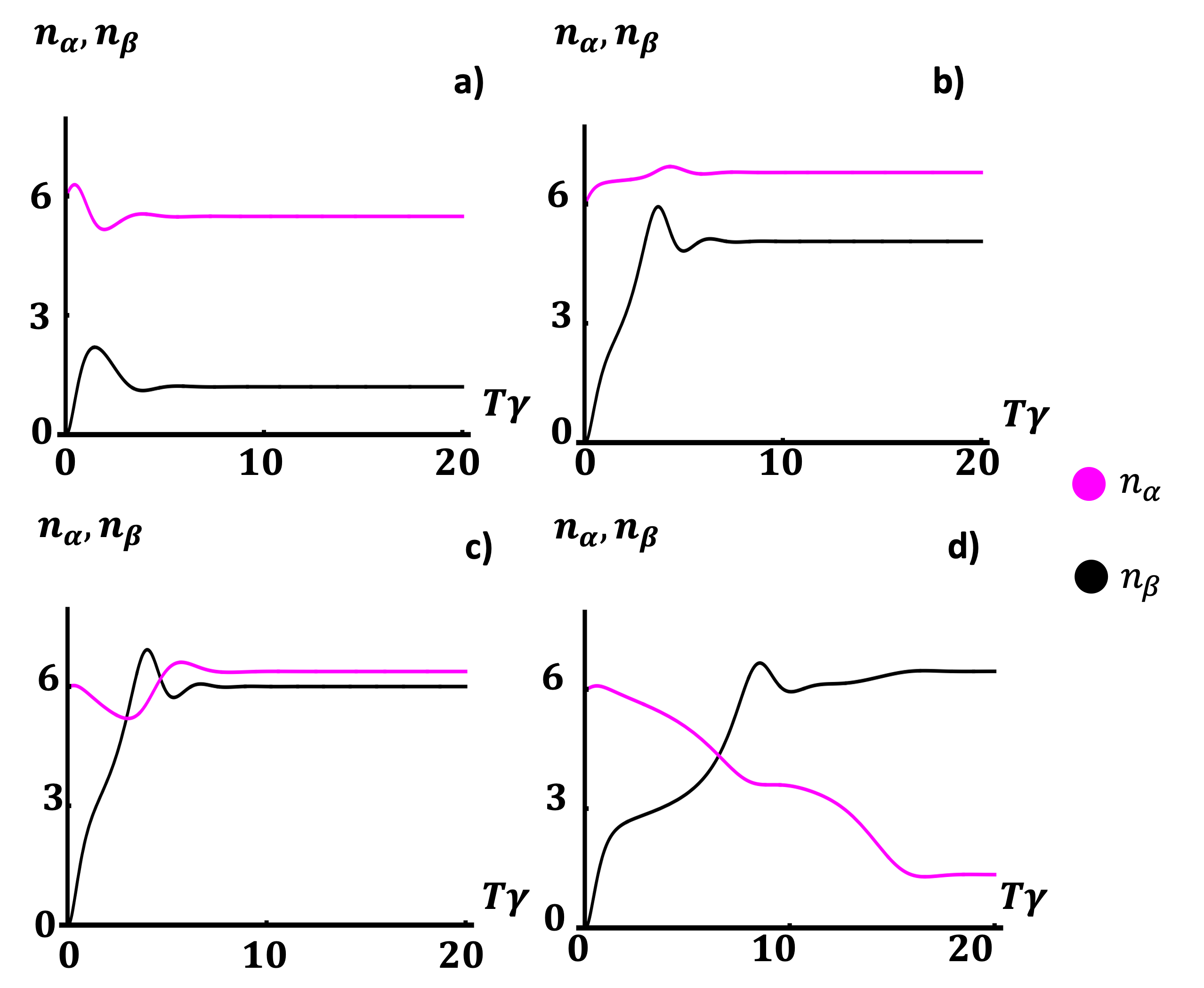}
    \caption{\justifying Long time dynamics driven-dissipative two model resonator obtained from semi-classically solving Eq.~(\ref{eq:QMaster}) with different vales of self Kerr term U and drive amplitude $F_{in}$. (a) Semiclassical steady state mode population for $U_1=0.5$ and $U_2 = 0.7$. (b) Reversed conditions of (a) with $U_1 = 0.7$ and $U_2 = 0.5$. (c) Analysis on effect of different drives with $F_1 = 2.5$ (purple) and $F_2 = 3.0$ (black) on the respective modes. Due to the presence of cross-Kerr term, the steady state almost collapse regardless how each mode is individually driven. The modes have different self Kerr terms with $U_1 = 0.6$ and $U_2 = 0.5$(d) Same parameters as in (c) but introduce $Im[J] = 0.072$, we observe again the same splitting as in ~\autoref{fig:steady-state}(c), indicating there are other parameter configurations to sense TRSB. Other parameters used: $\text{Re}[J] = 0.1$, $\kappa = 1$, $\Delta = -3.5$, $V = 0.1$and $\gamma = 2$. All parameters are given in units of dissipation $\gamma$. }
    \label{fig:diffU and F}
\end{figure}

In ~\autoref{fig:diffU and F} (a) we plot the steady-state and observe that the steady state does not collapse and the cross Kerr does not affect the system much. When inverting the self-Kerr terms in ~\autoref{fig:diffU and F} (b) that is $U_1 = 0.7$ and $U_2 = 0.5$, the situation changes drastically as with the presence of V, the steady states fundamentally get altered to become much closer. The system will remain capable of sensing TRSB signatures as long as the difference in self-Kerr terms is not too drastic and depends on the initial state conditions to which our system is subject. 

\begin{figure}[h]
    \centering
    \includegraphics[width=0.75\linewidth]{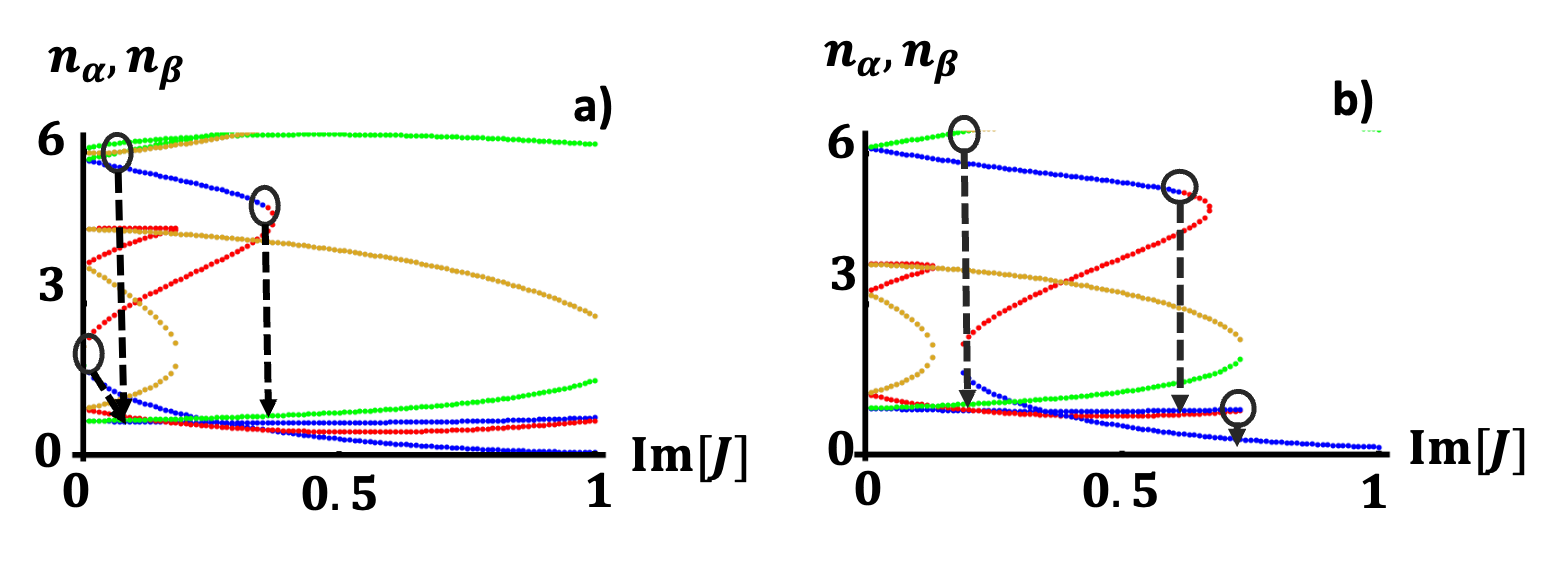}
    \caption{\justifying 2D Cross Section of 3D Visualization of mode population $n_\alpha$ and $n_\beta$ with Fig. 5 (a) with drive power $F_{in} = 2.7$ and Fig 5. (b) with driver power $F_{in} = 3.0$.}
    \label{fig:2D TRSB}
\end{figure}

When we investigate the effect of different drive terms in ~\autoref{fig:diffU and F} (c) that is $F_1 = 3$ and $F_2 = 2.5$ in the presence of $V=0.1$ and with $U_1 = 0.5$ and $U_2 = 0.6$ we observe that the system still collapses to almost the same steady state version. Interestingly, when having $Im[J] \neq 0$ as in ~\autoref{fig:diffU and F} (d) the steady state will be drastically lifted from its collapse and we can again probe the system for TRSB. These different variations show the richness of parameter space that can be probed with the ring resonator to sense TRSB signatures. Inherently, determining which parameter configurations allow for the highest sensitivity is a strenuous process and thus the analysis in our paper is mostly focused on symmetric conditions of self-Kerr $U$ and drive amplitude terms $F_{in}$.
We look at cross-sections of Fig. ~\ref{fig:classical} (c) with Fig. ~\ref{fig:2D TRSB} (a) and (b) for $F_{in}=2.5$ and $F_{in}=3$, respectively. 

We observe the asymmetric transitions of mode configuration in Fig. ~\ref{fig:2D TRSB} (a) at low $\text{Im}[J]$, whereas for Fig. ~\ref{fig:2D TRSB} (b) such asymmetric transitions appear weaker and for values of $\text{Im}[J]$ significantly larger at rates larger than most probe responses. This analysis underlines the ability of our device to sense TRSB stemming from its nonreciprocal nature of the modes and its interactions. \\

\section{Additional plots on hysteresis and photon number distribution for quantum analysis} \label{additional plots quantum}

\begin{figure}[h]
    \centering
    \includegraphics[width=0.5\linewidth]{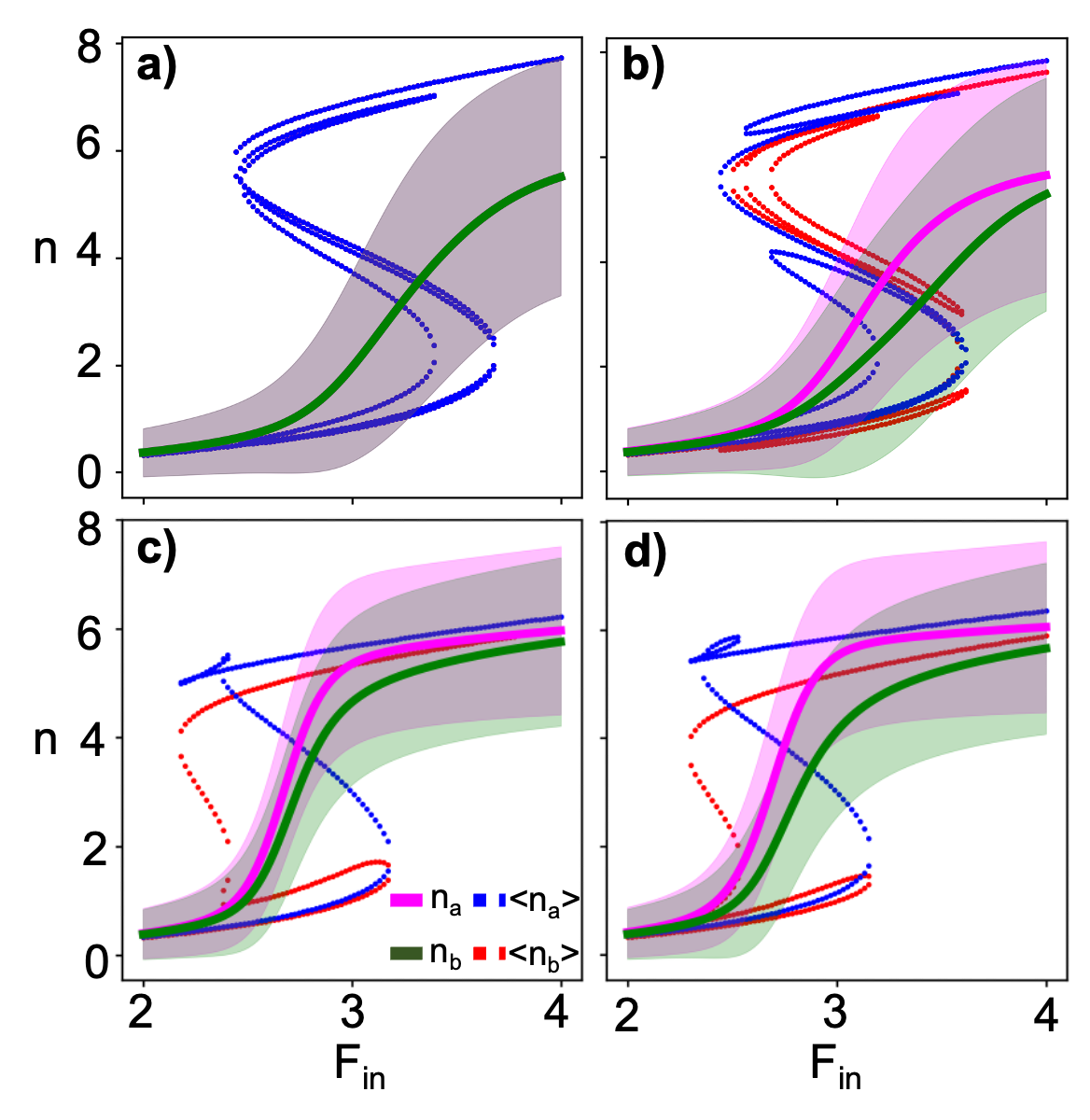}
    \caption{\justifying Comparison of semiclassical hysteresis curves and quantum steady-state solutions for different cross-Kerr interactions V and TRSB strengths $Im[J]$. Solid lines show semiclassical solutions for the photon populations $n_a$ and $n_a$, while markers indicate quantum steady-state expectation values $\langle n_a\rangle$ and $\langle n_b\rangle$. Shaded regions denote the variance associated with quantum fluctuations. Panels (a) show results without cross-Kerr interactions ($V=0$) and $Im[J]=0$. Panel (b) shows the corresponding hysteresis curve for $V=0$ and $Im[J]=0.1$, while panels (c–d) show the corresponding behavior for finite cross-Kerr coupling $V=0.2$ with (c) $Im[J]=0.1$ and (d) $Im[J]=0.2$. The presence of cross-Kerr interactions reduces fluctuations, enhances agreement between semiclassical and quantum results, and sharpens the separation of metastable branches, demonstrating the role of nonlinear interactions in stabilizing the sensing response.}
    \label{fig:quantum_extra}
\end{figure}

We provide additional hysteresis plots to compare to Fig.~\ref{fig:quantum}f to highlight the role of the cross Kerr term V and its effect on $Im[J]$. Fig.~\ref{fig:quantum_extra} plots the hysteresis curves for cross Kerr terms $V=0$ for $Im[J] = 0$ (a) and $Im[J] = 0.1$ (b) and $V=0.2$ for (c) $Im[J] =0.1$ and (d) $Im[J]=0.2$. 

We observe for Fig.~\ref{fig:quantum_extra} a that for no cross-Kerr or imaginary coupling term the semi-classical solutions collapses to a single mode solution as does quantum curve. The resulting noise (grey background) also converges to a unimodal state. Once the imaginary term $Im[J]$ is introduced in Fig.\ref{fig:quantum_extra}b a splitting of the steady states emerges, corresponding to the onset of asymmetric mode populations. However, in the absence of cross-Kerr interactions this splitting remains relatively weak and the quantum solutions exhibit broad fluctuations, indicating that switching between configurations remains significant.

When the cross-Kerr term is introduced, as shown in Fig.~\ref{fig:quantum_extra}(c) for $V=0.2$ and $Im[J]=0.1$, the fluctuations are noticeably reduced and the quantum solutions more closely follow the semiclassical branches. The nonlinear interaction effectively sharpens the potential landscape, increasing the residence time near metastable configurations and thereby enhancing the distinguishability of the branches. At the same time, the competition between the cross-Kerr interaction and the TRSB perturbation can partially suppress the population imbalance, leading to smaller separation of the quantum curves at fixed $Im[J]$.

Increasing the TRSB strength further, as shown in Fig.~\ref{fig:quantum_extra}(d) for $Im[J]=0.2$, restores a stronger separation between the branches, while the fluctuations increase correspondingly. This behavior illustrates the trade-off between nonlinear stabilization and signal strength: larger cross-Kerr interactions reduce noise and improve agreement with semiclassical predictions, whereas stronger TRSB perturbations increase the measurable population imbalance but also enhance fluctuations.

\section{Definition of Probability Density Function}
\label{PDF}

The probability density function (PDF) defines the likelihood of a continuous random variable falling within a particular range of values. For our particular case in Fig.~\ref{fig:quantum}, we employ a normal distribution where generally the PDF is defined as:

\begin{equation}
    f(x) = \frac{1}{\sqrt{2\pi \sigma^2}}e^{-\frac{(x-\mu)}{2\sigma^2}}
\end{equation}

where $x$ is the support,$\mu$ is the mean and $\sigma^2$ the variance. For the photon number histograms in Fig~\ref{fig:quantum}, the mean and variance are defined as:

\begin{align}
    x &= N \nonumber \\ 
    \mu &= \sum_{i\in\{a,b\}} P_i(n) * N \\
    \sigma^2 &= Var{\Delta n} = \langle (n_a - n_b)^2 \rangle - \langle (n_a - n_b) \rangle ^2 \nonumber
\end{align}

where $P_i (n) = diag[Tr_i \rho]$ is the photon occupation probability and $N \in [0,12]$ the number of photon states probed. $n_a$ and $n_b$ represent the photon distributions for mode a and mode b. 

For the time averaged case where $\sigma_{avg} (\tau) \approx \sigma_{shot}/\sqrt{\kappa \tau}$, where $\sigma_{shot} = \sqrt{{\rm var}(\Delta n)}$, the variance in the PDF is replaced with the time averaged component. 

\end{document}